# Two-wire Terahertz Fibers with Porous Dielectric Support

**Andrey Markov, and Maksim Skorobogatiy\***

*Department of Engineering Physics, École Polytechnique de Montréal, Québec, Canada*
*\*maksim.skorobogatiy@polymtl.ca*

**Abstract:** A novel plasmonic THz fiber is described that features two metallic wires that are held in place by the porous dielectric cladding functioning as a mechanical support. This design is more convenient for practical applications than a classic two metal wire THz waveguide as it allows direct manipulations of the fiber without the risk of perturbing its core-guided mode. Not surprisingly, optical properties of such fibers are inferior to those of a classic two-wire waveguide due to the presence of lossy dielectric near an inter-wire gap. At the same time, composite fibers outperform porous fibers of the same geometry both in bandwidth of operation and in lower dispersion. Finally, by increasing cladding porosity one can consistently improve optical properties of the composite fibers.

## 1. Introduction

The main complexity in designing terahertz waveguides is the fact that almost all materials are highly absorbent in the terahertz region [1]. Since the lowest absorption loss occurs in dry air, an efficient waveguide design must maximize the fraction of power guided in the air. Different types of THz waveguides and fibers have been proposed based on this concept. The simplest subwavelength fiber [2-4] features dielectric core that is much smaller than the wavelength of guided light. As a result, a high fraction of modal power is guided outside of the lossy material and in the low-loss gaseous cladding. Another type of the low-loss fibers includes fibers featuring porous core region with the size of the individual pores much smaller than the wavelength of light [2, 3]. Consequently, guided light has a strong presence in the low-loss gas-filled pores inside the core. High modal confinement in the core makes such fibers less prone to bending losses and less sensitive to the environment compared to the simple rod-in-the-air subwavelength fibers [2, 5]. Another important type of the low-loss fibers includes fibers featuring gas-filled core surrounded by a structured cladding serving as a reflector. The main challenge in the design of such fibers is to ensure high reflection at the core-cladding interface. Different hollow-core structures have been investigated including metalized bores [5-7], periodic dielectric multilayers [8], as well as thin-walled dielectric pipes [9-11]. Among other types of the THz waveguides, we note parallel plate waveguides [12] and slit waveguides [13], which are known for their low losses and strong confinement.

Metal wire, can be used to transport terahertz pulses with virtually no dispersion and low attenuation [14]. However, it is difficult to realise efficient coupling into the wire-bound mode. This is because the fundamental mode of a single wire is radially polarized, thus, commonly used photoconductive antennas that tend to produce linearly polarized THz light cannot be used directly for efficient excitation of the wire mode. In addition, high bending losses of a single wire waveguide limit its practical applications. Thus, even a slight bending of the wire can lead to considerable increase in the modal transmission loss, e.g. from 0.03 cm$^{-1}$ for a straight wire to 0.05 cm$^{-1}$ for a slightly bend one (bending radius of 90 cm, [15]).

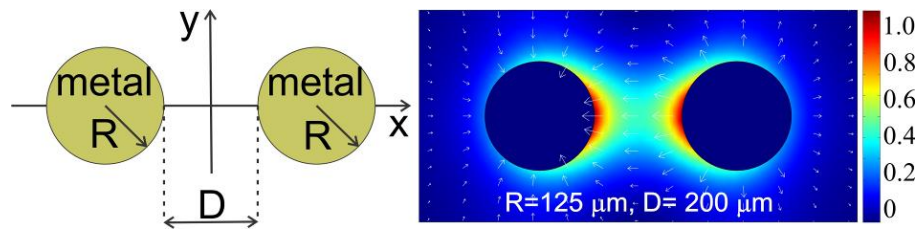

Fig. 1. a) Schematic of a classic two-wire waveguide. b) Longitudinal flux distribution for the TEM mode of a two-wire waveguide. Arrows show vectorial distribution of the corresponding transverse electric field.

The two-wire waveguides [16] combine low loss performance and high excitation efficiency. Particularly, field distribution in the fundamental transverse-electromagnetic (TEM) mode of a two-wire waveguide (see Fig. 1) has the same symmetry as that of an electromagnetic wave emitted by a simple THz dipole source (photoconductive antenna) when the wave is polarised along the line joining the two wires. Thus, one expects high excitation efficiencies of the fundamental mode of a two-wire waveguide when using standard dipole terahertz sources. Moreover, efficient confinement of the modal energy between the two wires, as opposed to highly delocalised Sommerfeld wave on a single wire, makes two-wire waveguides less prone to the bending losses. Thus, for the same bending radius, bending loss of a single wire waveguide can be more than 5 times higher than that of a two-wire waveguide [16]. Additionally, absorption losses, and group velocity dispersion (GVD) of the fundamental mode of a two-wire waveguide are extremely low (see Fig. 2).

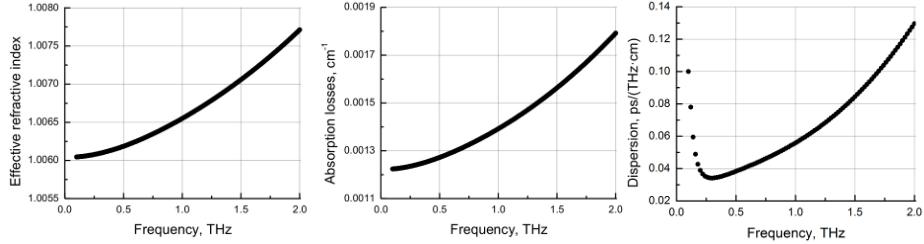

Fig. 2. Effective refractive index, absorption losses and group velocity dispersion of the fundamental mode of a two metal wire waveguide shown in Fig. 1.

While having outstanding optical properties, a classic two-wire waveguide is inconvenient in practical applications. Thus, in a typical experiment the two wires have to be aligned and kept straight and parallel to each other with high precision. This requires bulky holders and cumbersome coupling setups. Moreover, the fiber core is not encapsulated into a protective cladding, thus leaving the core (space between wires) exposed to the environment.

In this paper, we explore several practical designs of the two-wire waveguides for THz guidance. Particularly, we aim at designing two-wire waveguides that are more convenient to manipulate mechanically without disturbing the structure of the guided mode, while retaining as much as possible the outstanding optical properties of a classic two-wire waveguide. The proposed waveguide structure consists of a porous polyethylene fiber and two metal wires that are positioned inside of a porous cladding. In its simplest implementation, a polyethylene fiber has three interconnected circular holes, among which the central hole is empty and the two peripheral holes are filled with the metal wires. Designs that are more complex feature a larger number of holes or a web of thin bridges in the air that are arranged within the fiber outer jacket. The metal wires are placed into two fiber holes with an inter-hole separation comparable to the wire diameter. In all these designs, a significant fraction of the modal power is propagating in the air-filled porous core, which reduces the modal absorption losses as well as the modal group velocity dispersion. Finally, we believe that fibers described in this paper can be useful not only for low-loss THz wave delivery but also for sensing of biological and chemical specimens in the terahertz region by placing the recognition elements directly into the fiber microstructure similarly to what has been recently demonstrated in [17].

## 2. Classic two-wire waveguide

Before we detail the novel composite fibers we would like to address the issue of coupling efficiency into a classic two-wire waveguide. From our experiments, we observe that this coupling is a sensitive function of the excitation wavelength. As it is well known, the coupling efficiency into a classic two-wire waveguide achieves its maximal value at the wavelength that is comparable to the inter-wire separation, while the coupling efficiency stays relatively low for the wavelengths that are significantly smaller or larger than the optimal one. Ultimately, it is the frequency dependent coupling efficiency that limits practically usable bandwidth in such waveguides. Moreover, coupling efficiency into two-wire waveguides depends strongly on several geometric parameters such as position of the THz beam focal point, the width of the wires and the distance between them. For the completeness of presentation, we first study numerically the influence of each of these parameters on the waveguide excitation efficiency.

Modes of the waveguides and fibers studied in this work were computed using COMSOL Multiphysics FEM mode solver. The frequency dependent relative permittivity and conductivity of metal in the THz spectral range is modeled using the Drude formula:

$$\varepsilon(\omega) = 1 - \frac{\omega_p^2}{\omega^2 + i\omega\Gamma} \underset{THz}{\approx} -\frac{\omega_p^2}{\Gamma^2} + i \cdot \frac{\sigma}{\omega \cdot \varepsilon_0} \quad ; \quad \sigma \underset{THz}{\approx} \frac{\varepsilon_0 \omega_p^2}{\Gamma}, \quad (1)$$

where $\varepsilon_0$ is the free-space permittivity, $\omega$ is the angular frequency, $\omega_p$ is the angular plasma frequency, and $\Gamma$ is the electron scattering rate. In our simulations we assume copper as a metal, and we use the following values of the parameters: $\omega_p = 2\pi \cdot 1.969 \cdot 10^{15}\ Hz$, $\Gamma = 2\pi \cdot 4.775 \cdot 10^{12}\ Hz$ (derived from [18-20]). In the THz frequency range the Drude model is especially simple as it predicts frequency independent real part of the relative permittivity $\varepsilon_r = -1.7 \cdot 10^5$, and frequency independent conductivity $\sigma = 4.5 \cdot 10^7\ S/m$.

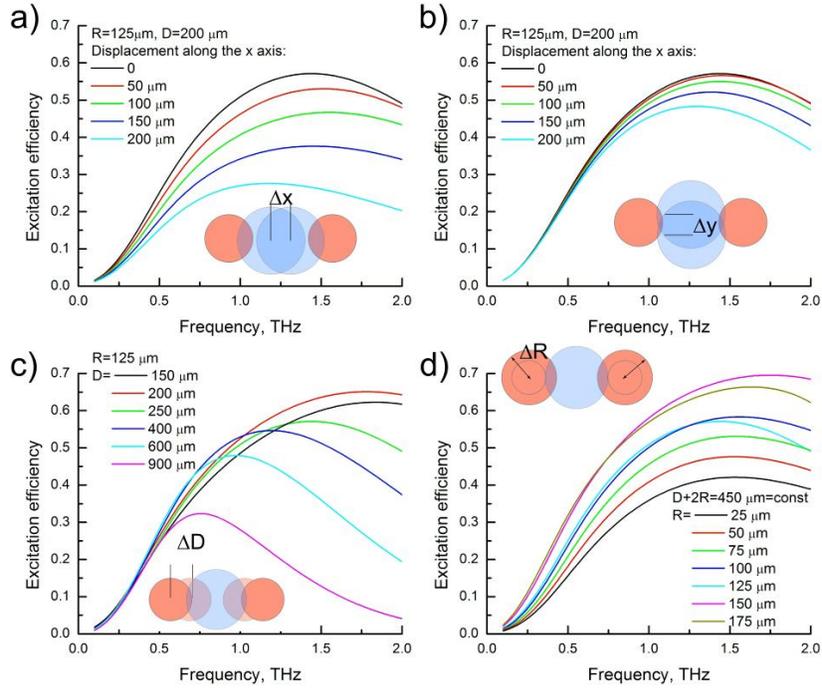

Fig. 3. Excitation efficiency of the fundamental mode of a classic two-wire waveguide using Gaussian beam as an excitation source. Dependence of the excitation efficiency on various geometrical parameters, such as: a) displacement along the x axis from the core center; b) displacement along the y axis from the core center; c) inter-wire gap size; d) wire radius.

Coupling efficiency into a two-wire waveguide (defined as a fraction of the coupled power to the power in the excitation beam) was computed similarly to our prior work [3] assuming as an excitation source a Gaussian THz beam with a frequency-dependent beam diameter $d_0 \approx 2.5 \cdot \lambda$. Efficient coupling of the Gaussian beam into the fundamental mode of a two-wire waveguide requires a linear polarization of the THz beam with electric field directed along the line connecting the two wires. As modal power in a two-wire waveguide is mainly confined between the wires, we expect the highest coupling efficiency when the Gaussian beam diameter is comparable to the size of the inter-wire gap.

In Fig. 3, we present excitation efficiency of the fundamental mode of a two-wire waveguide using Gaussian beam as a source. We consider effect of the various geometric parameters on the coupling efficiency, including misalignment of the THz beam relative to the position of the wires, as well as choice of the wire radius and size of the inter-wire gap.

First, we study excitation of the fundamental TEM mode of a two-wire waveguide using Gaussian beam with the focal point displaced from the mid-point between the two wires (the point of maximal coupling efficiency). From Fig. 3(a) we see that displacement of the beam focal point along the OX axis by 100 μm leads to 20% decrease in the TEM excitation efficiency in the whole THz range, while displacement by 200 μm leads to 50% drop in the TEM excitation efficiency. The effect of displacement of the Gaussian beam focal point along the OY axis is much less pronounced (see Fig. 3(b)). Thus, displacement of the beam focal point by 100 μm from the waveguide center leads to only 5% decrease in the TEM excitation efficiency. Remembering that the inter-wire gap size is 200 μm, we conclude that coupling efficiency is only moderately sensitive to the errors in the positioning of the Gaussian beam.

We now examine excitation of the TEM mode using a perfectly centered Gaussian beam while varying distance between the two wires. From Fig. 3(c) we find that the TEM mode is excited most efficiently when the Gaussian beam waist is comparable to the size of the inter-wire gap. The excitation efficiency remains high 40 – 70% in the whole THz frequency range 0.5 – 2.0 THz when the inter-wire distance is 100 – 400 μm. Finally, for a fixed 450 μm center-to-center distance between the two wires, in Fig. 3(d) we plot excitation efficiency as a function of the wire radii. For small values of the radius the excitation efficiency increases with the radius until reaching its maximal value at R= 150 μm. Further increase of the wire radius leads to a monotonous decrease of the coupling efficiency.

## 3. Composite fiber featuring two metal wires in a three-hole porous cladding

Presently, there are no fibers operating in the THz frequency range that are low-loss, low-dispersion, and broadband at the same time. The two-wire waveguides open a possibility for making practical THz fibers since they offer low-loss, low-dispersion transmission and good coupling efficiency to standard photoconductive antennas. Strong optical performance of the two-wire waveguides is a direct consequence of the modal confinement predominantly in the low-loss, low-dispersion dry gas between the two wires. At the same time, practical two-wire waveguides require some form of a robust mechanical support or overcladding in order to hold the metallic wires straight and parallel to each other, to optically separate the core region from the environment, and to provide encapsulation for the dry gas in the vicinity of the wires. We propose using porous polyethylene fibers that were demonstrated to exhibit excellent guiding characteristics at low frequencies [21-23] to provide a convenient packaging solution for the two-wire waveguides. In their simplest implementation such fibers can contain three adjacent holes with the two of them occupied by the metal wires, while the central one kept empty and used to guide a larger portion of the THz radiation (see Fig. 4).

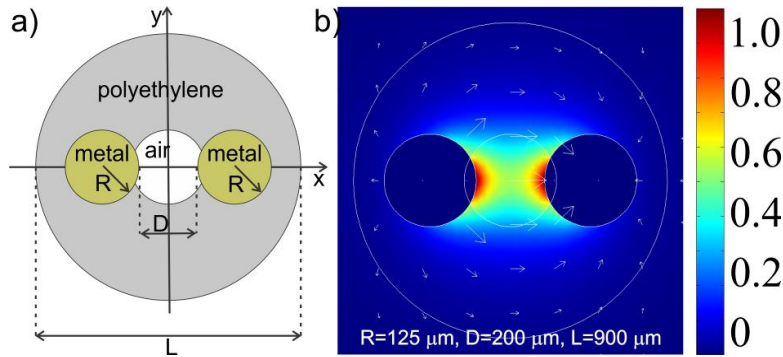

Fig. 4. a) Schematic of a composite fiber featuring two metal wires in a three-hole cladding. b) Longitudinal flux distribution of a typical guided mode presents a mixture of the plasmonic mode guided by the metal wires and a TIR mode guided by the porous fiber cladding.

Naturally, the wave guiding in such fibers is most efficient for the light polarized parallel to the line connecting the wires. A typical modal pattern represents the mixture of a plasmonic mode guided by the two wires, and a TIR (total internal reflection) mode guided by the fiber plastic cladding (see Fig. 4). For the perpendicularly polarized light, efficient excitation of the plasmonic mode is not possible, and the fiber guides largely as a porous TIR fiber studied earlier [21]. In this work, we therefore concentrate only on the parallel polarisation.

The modal dispersion relations, the absorption losses, and the excitation efficiencies of the various modes of a composite two-wire fiber are presented in Fig. 5 in black color. In our simulations we used 1.514 as a frequency independent refractive index of polyethylene with frequency dependent material loss $\alpha \left[ cm^{-1} \right] = 0.28 \cdot \nu^2$ (operation frequency $\nu$ is in THz).

The presence of porous polyethylene cladding significantly complicates the fiber modal structure. In order to distinguish predominantly plasmonic modes from the modes of a porous cladding, in Fig. 5 we also present (in red color) optical properties of the modes of a porous cladding alone (no wires). Note that among the modes of a composite fiber, there is one that has no corresponding analogue among the modes of a porous cladding, which is especially evident at lower frequencies <0.25 THz. In what follows, we call this mode a "fundamental plasmonic mode." Note that addition of porous cladding shifts the frequency of maximal excitation efficiency of the fundamental plasmonic mode from 1.5 THz in a classic two-wire waveguide to a much lower frequency of ~0.4THz in a composite fiber.

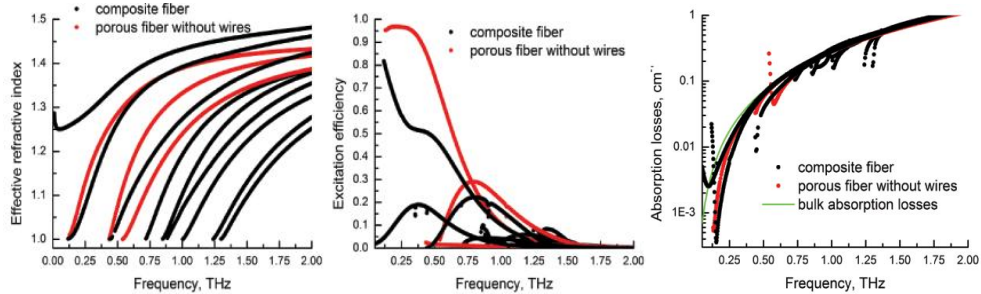

Fig. 5. Black color: the effective refractive indices, absorption losses and excitation efficiencies for the various modes of a composite two-wire fiber shown in Fig. 4. Red color: various optical properties of the modes of a corresponding porous cladding (no metal wires).

We now consider in more details the modes of a 3-hole composite fiber. In Fig. 6 we present again the effective refractive indices, the absorption losses and the excitation efficiencies of the composite fiber modes, however, this time we concentrate on the sub – 1 THz frequency range. Moreover, solid colors in Fig. 6 define frequency ranges where modal excitation efficiency is higher than 5%, while dashed curves define spectral regions with less than 5% coupling efficiency from the Gaussian beam. In these simulations, we assume that the fiber center coincides with the focal point of a Gaussian beam.

First, we concentrate on the fundamental plasmonic mode of a composite fiber. Optical properties of this mode are presented in Fig. 6 in blue color. Coupling efficiency into the fundamental plasmonic mode varies between 5% – 20% in the 0.13 – 0.79 THz frequency range. As shown in Fig. 7, at lower frequencies 0.1 – 0.4 THz, electric field of the mode is mostly oriented along the line connecting the two wires, and the corresponding modal power is concentrated in the air–filled central hole. This results in modal losses that can be considerably smaller than the bulk absorption loss of polyethylene. At higher frequencies (>0.4 THz) the modal power is displaced away from the central air hole (see Fig. 7) and into the polyethylene cladding surrounding the wires. Consequently, at higher frequencies absorption loss of the fundamental plasmonic mode approaches the bulk absorption loss of polyethylene. Furthermore, from the modal dispersion relation presented in Fig. 6 it follows that the group velocity dispersion of the fundamental plasmonic mode is ~2-3 ps/(THz·cm), which is much higher than that of a fundamental mode of a classic two-wire waveguide (~0.1

THz/(ps·cm)). The reason for the relatively high value of the group velocity dispersion of the fundamental plasmonic mode of a composite fiber is in the rapid change of the modal confinement pattern in the 0.1-0.8 THz range from mostly air at lower frequencies to both air and polyethylene at higher frequencies. At the same time, at lower frequencies (<0.3 THz), fundamental plasmonic mode of a composite fiber has a much smaller group velocity dispersion when compared to the ~10 ps/(THz·cm) dispersion of the fundamental mode of the corresponding porous fiber (without metal wires).

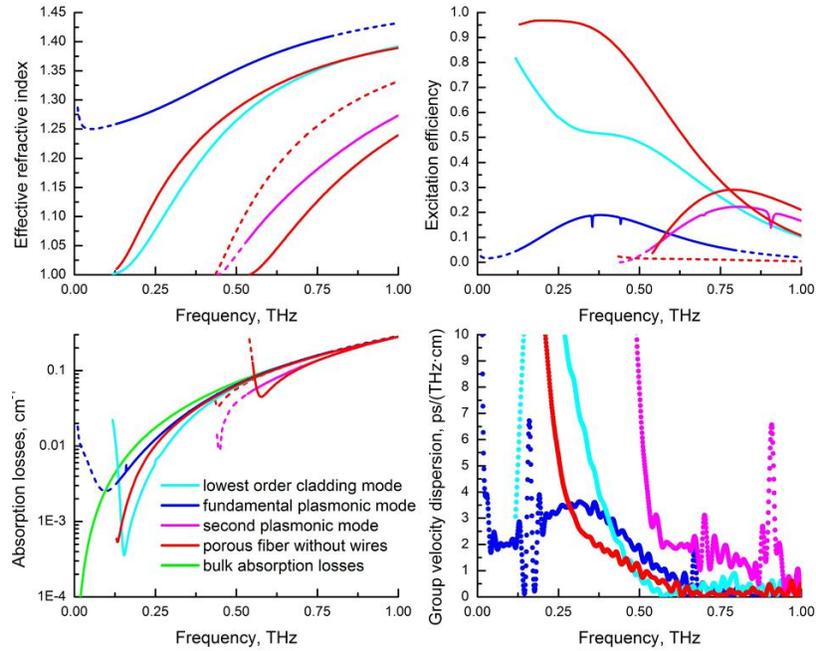

Fig. 6. Effective refractive indices, excitation efficiencies, absorption losses and group velocity dispersion for the various modes of a composite two-wire fiber shown in Fig. 4.

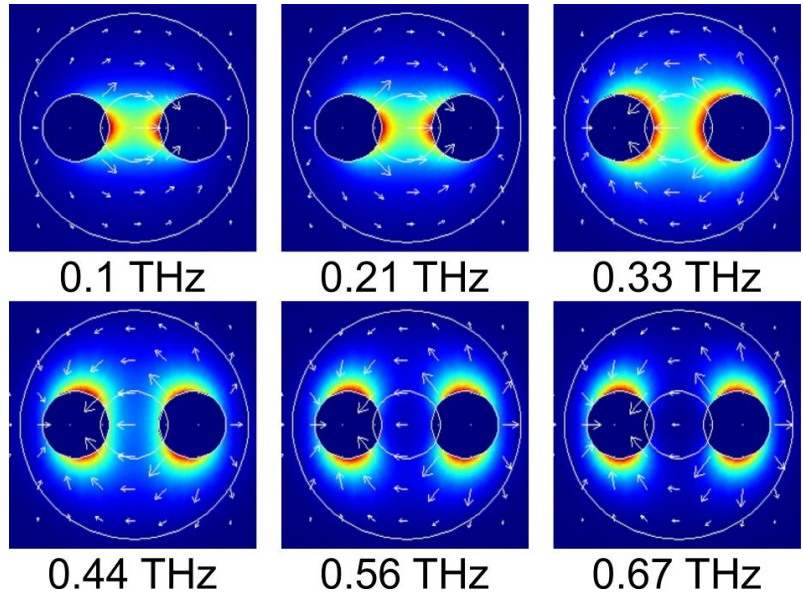

Fig. 7. Longitudinal flux distribution of the fundamental plasmonic mode of a composite fiber at various operation frequencies.

We now consider the lowest order cladding mode of a composite fiber in the 0.12 – 1.00 THz frequency range (cyan color in Fig. 6). At lower frequencies (<0.3 THz) when operation wavelength is larger or comparable to the fiber size, the lowest order cladding mode has a strong presence outside of the fiber (see Fig. 8). At higher frequencies, the modal fields tend to confine inside of the fiber cladding, thus resulting in the mode absorption losses similar to the bulk absorption losses of the cladding material. Note that the lowest order cladding mode of a composite fiber is in fact a hybrid mode that has a significant plasmon contribution. For this particular mode, the plasmon is propagating at the plastic/metal interface with almost no energy found in the central air hole.

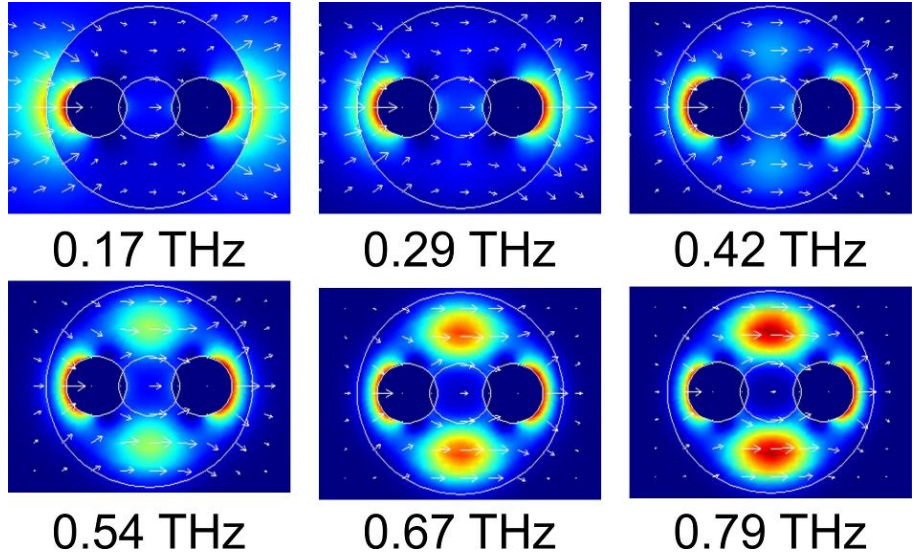

Fig. 8. Longitudinal flux distribution of the lowest order cladding mode of a composite fiber at various frequencies.

It is interesting to compare cladding modes of a composite fiber with the modes of a porous fiber of the same cross section, however, without the metallic wires. Optical properties of the modes of a porous fiber are presented in Fig. 6 in red color. We note that in the broad frequency range 0.13 – 1.0 THz, optical properties of the fundamental mode of a porous fiber are quite similar to those of the lowest order cladding mode of a composite fiber. At the same time, the corresponding field distributions are somewhat different at low frequencies. Thus, below 0.5 THz (see, for example, Fig. 9, 0.17THz), field distribution of the fundamental mode of a porous fiber shows significant presence in the air cladding outside of the fiber, as well as in the air holes inside of a plastic cladding. In contrast, lowest order cladding mode of a composite fiber (see Fig. 8, 0.17THz), shows no field presence in the central air hole between the two metal wires, while having significant field concentration in the air region outside of the fiber. At higher frequencies (Figs. 8, 9, >0.5THz) field distributions of the two modes become very similar to each other and feature strong light localisation in the plastic fiber cladding. Consequently, optical properties of the two modes also become very similar at higher frequencies.

Finally, we consider the second order cladding mode of a composite fiber, which is presented in Fig. 6 in magenta color in the frequency range of 0.54 – 1.0 THz. From the corresponding field distributions showed in Fig. 10 it follows that the second order cladding mode is, in fact, a hybrid mode that has a significant plasmon contribution. Moreover, at lower frequencies (0.5 – 0.7 THz) the plasmon is propagating at the air/metal interface with a significant amount of energy concentrated in the central air hole between the two wires. In what follows we call this mode the second plasmonic mode as it can be used for low loss guidance of THz light within the central air hole of a composite fiber.

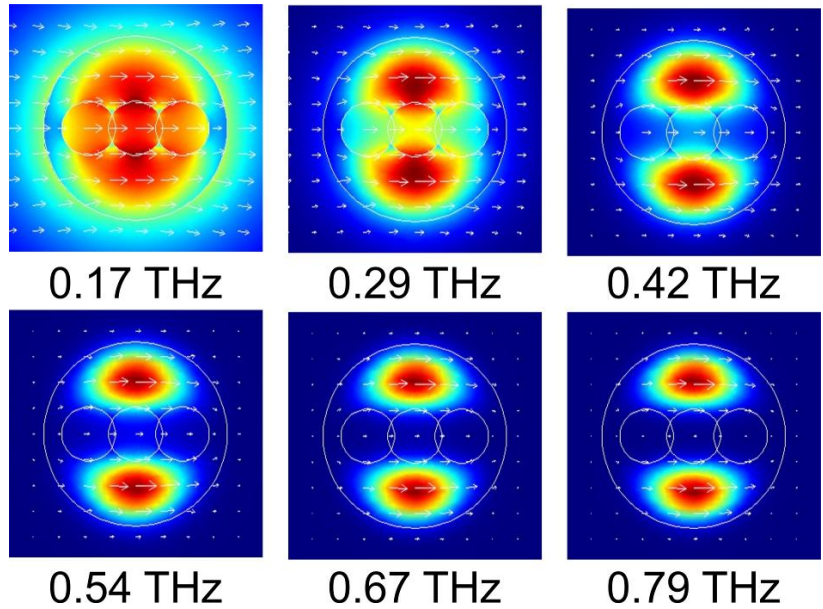

Fig. 9. Longitudinal flux distribution of the fundamental mode of a porous fiber (same cross section as in Fig. 4, however, without metal wires).

At higher frequencies (>0.7 THz), the second plasmon mode leaves the central air hole, while localising in the vicinity of the metal/plastic/air junctions (see, for example, Fig. 10, 0.79 THz). This results in significant energy transfer into the cladding, and, consequently, modal absorption losses of this mode become comparable to the bulk absorption losses of a polyethylene cladding. In the 0.5 – 0.9 THz spectral range, the second order plasmonic mode has relatively low group velocity dispersion ~1-2 ps/(THz·cm), which is even smaller than that of the fundamental plasmonic mode.

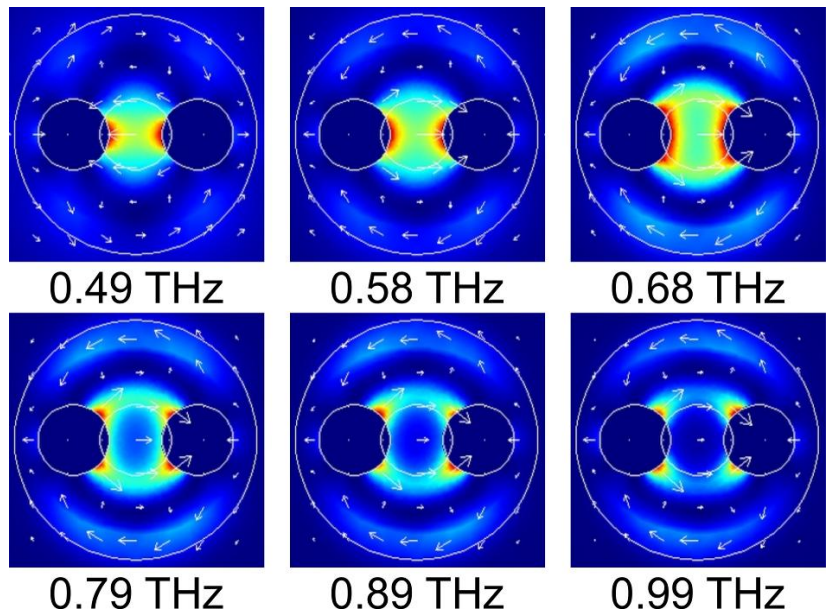

Fig. 10. Longitudinal flux distribution of the second plasmonic mode of a composite fiber.

It is now timely to highlight the main differences in the optical properties of the modes of a two-wire composite fiber and the modes of a corresponding porous fiber without metal wires. Firstly, from Fig. 6 we note that the fundamental plasmonic mode of a composite fiber extends into very low frequencies (<0.1THz), while being well confined within the fiber (see Fig. 7). This can be of advantage when compared to the fundamental mode of a porous fiber, which at low frequencies is highly delocalised in the air cladding outside of the fiber (see Fig. 9). In practical terms it means that at very low frequencies, the fundamental plasmonic mode of a composite fiber is still suitable for guiding THz light due to its strong confinement in the core and, consequently, low sensitivity to bending, imperfections on the fiber surface, as well as perturbations in the environment. At the same time, at very low frequencies, fundamental mode of a porous fiber is largely found outside of the fiber core, which makes it highly sensitive to perturbations in the environment, as well as bending and fiber surface quality.

Secondly, cladding modes of a composite fiber tend to have lower absorption losses than the corresponding modes of a porous fiber without wires. This is mostly due to enhanced expulsion of the modal fields from the lossy plastic cladding near the metal wires.

Finally, we conclude that the fundamental mode of a classic two-wire waveguide can still be recognised in the modal structure of a composite two-wire fiber. In this case, however, the pure plasmonic mode is hybridised with the modes of a plastic cladding, thus, leading to increased modal losses and increased modal group velocity dispersion compared to those of the fundamental mode of a two-wire waveguide. At the same time, predominantly "plasmonic" modes of a composite fiber, while exhibiting comparable absorption losses, possess significantly lower group velocity dispersion than the modes of a corresponding porous fiber. Therefore, under the restricted launch conditions, when mostly plasmonic modes of the composite fiber are excited, we believe that such fibers can outperform both in loss and in the effective group velocity dispersion the corresponding porous fibers without metal wires.

## 4. Composite fiber featuring two metal wires in a seven-hole porous cladding

In the previous section, we have observed that absorption losses of the modes of a three-hole composite fiber were quite smaller than the bulk absorption losses of a plastic cladding. This is a direct consequence of the modal localisation in the fiber central air hole and in the air region outside of the fiber. At the same time, losses and group velocity dispersion of the modes of a three-hole composite fiber were still much higher than those of a classic two-wire waveguide. This is due to the fact that many modes of a composite fiber bare strong resemblance to the modes of a porous three-hole plastic fiber without wires.

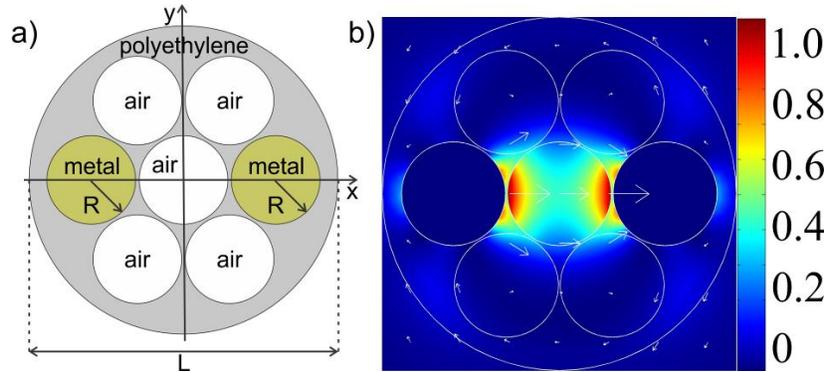

Fig. 11. a) Schematic of a seven-hole fiber with two metal wires. b) Longitudinal flux distribution for a typical guided mode of a composite fiber presents a mixture of the plasmonic mode guided by the metal wires and a TIR mode guided by the fiber cladding.

As we saw in our prior work [21-23], group velocity dispersion of the modes of a porous fiber can be quite large especially in the spectral region where modal guidance mechanism changes from low confinement (subwavelength guidance) to strong confinement in the dielectric cladding. As demonstrated in [24] this transition happens in the vicinity of the following characteristic frequency:

$$\nu_0 \approx \frac{c}{2\pi R_f \sqrt{(1-f)(\varepsilon_c - \varepsilon_a)}}, \quad (2)$$

where $c$ is the speed of light in vacuum, $R_f$ is the fiber radius, $\varepsilon_c$ and $\varepsilon_a$ are the relative permittivities of the cladding material and air respectively, and $f$ is the fiber porosity defined as a ratio of the total area taken by the holes to the total area of the fiber crossection. In order to keep the influence of plastic cladding to the minimum both in terms of the material absorption loss and in terms of the group velocity dispersion, we need to ensure that the operation frequency range of a two-wire waveguide corresponds to the subwavelength guidance regime of a porous cladding. For example, for a composite fiber of 1 mm diameter made of polyethylene, from (2) it follows that in order for the fiber cladding to operate in the low confinement (subwavelength) regime at 1 THz, porosities of 99.3% are required.

Although in practice, such high porosities are challenging to achieve, it is clear that increasing porosity of the composite fiber cladding should improve the fiber optical properties. To demonstrate this point we study a simple seven-hole fiber with air-holes placed in the vertices of an equilateral triangular lattice (see Fig. 11 (a)). The metal wires are placed in the two opposing holes of the fiber, thus forcing THz light to be guided predominantly in the central air hole (see Fig. 11 (b)). In order to decrease the number of cladding modes, thickness of the outer cladding has to be minimized. This thickness is defined as the smallest distance between the fiber outer surface and the boundary of the internal air holes. From our experience with plastic porous fibers and capillaries [11, 21], outer cladding thickness larger than 50 μm is sufficient to mechanically protect the inner structure of the fiber. In our calculations, the smallest bridge size between any two air holes is taken to be 10 μm, which can be readily realised in practice as demonstrated in our prior experiments with suspended core fibers [3]. The outer diameter of the seven-hole fiber is 870 μm, the wire diameter is the same as before and equal to 250 μm, and the air hole diameter is equal to the wire diameter.

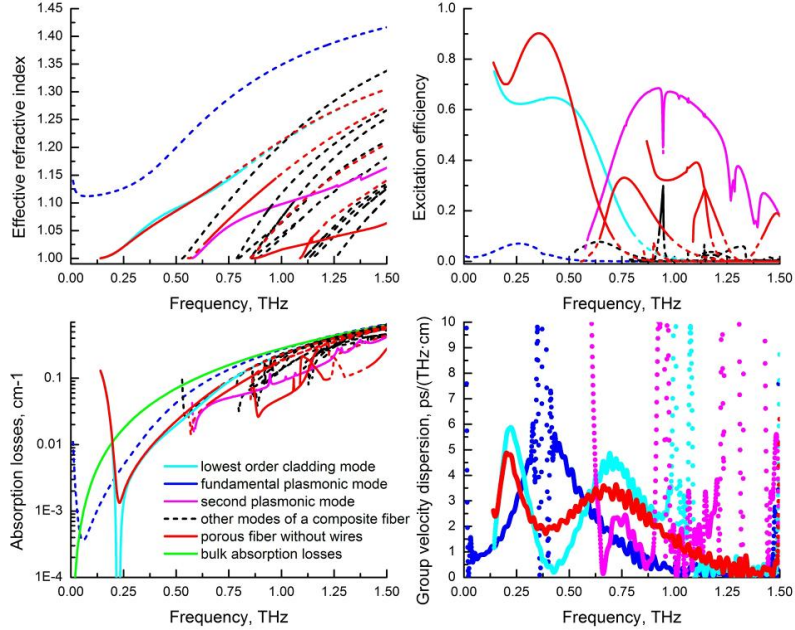

Fig. 12. Effective refractive indices, excitation efficiencies, absorption losses and group velocity dispersion for the various modes of a seven-hole composite fiber shown in Fig. 11. Solid lines define frequency ranges where the modal excitation efficiency is higher than 10%.

In Fig. 12 we present dispersion relations, absorption losses, coupling efficiencies, and group velocity dispersion of the modes of a seven-hole composite fiber shown in Fig. 11(a). First, we observe that at lower frequencies (<1.0 THz) modal structure of a seven-hole composite fiber is similar to that of a three-hole composite fiber (see Fig. 6). Thus, we distinguish the "fundamental plasmonic mode" (blue color in Fig. 12), which has a strong presence in the central air hole at frequencies below 0.4 THz (see Fig. 13(a)). Unfortunately, coupling to this mode is quite inefficient and is below 10% in the whole THz range. Next, we note that optical properties of the lowest order cladding mode (cyan color in Fig. 12) resemble closely those of the corresponding mode of a three-hole fiber. The major difference between the two modes is observed at lower frequencies <0.3 THz where modal field presence in the air regions is more pronounced in a seven-hole fiber due to higher porosity of its plastic cladding (compare Figs. 13(b), 8). Finally, when comparing optical properties of the second plasmonic mode of the two fibers, we note that excitation efficiency of this mode in a seven-hole fiber is relatively high (20-70%) in a broad spectral range 0.59 THz-1.7 THz, while the maximal excitation efficiency of this mode in a three-hole fiber is only 20% at 0.8 THz. Moreover, as seen from Figs. 13(c), 10, at higher frequencies, modal field of the second plasmonic mode of a seven-hole fiber has a much stronger presence in the central air-hole when compared to the three-hole fiber. This, again, can be explained by the higher porosity of the seven-hole fiber. Finally, from Fig. 12 we expect that even at higher frequencies ~1THz transmission losses of a seven-hole fiber should be significantly smaller (3-5 times) than the bulk absorption loss of a cladding material, while the fiber effective group velocity dispersion should be in the ~1-4 ps/(THz·cm) range.

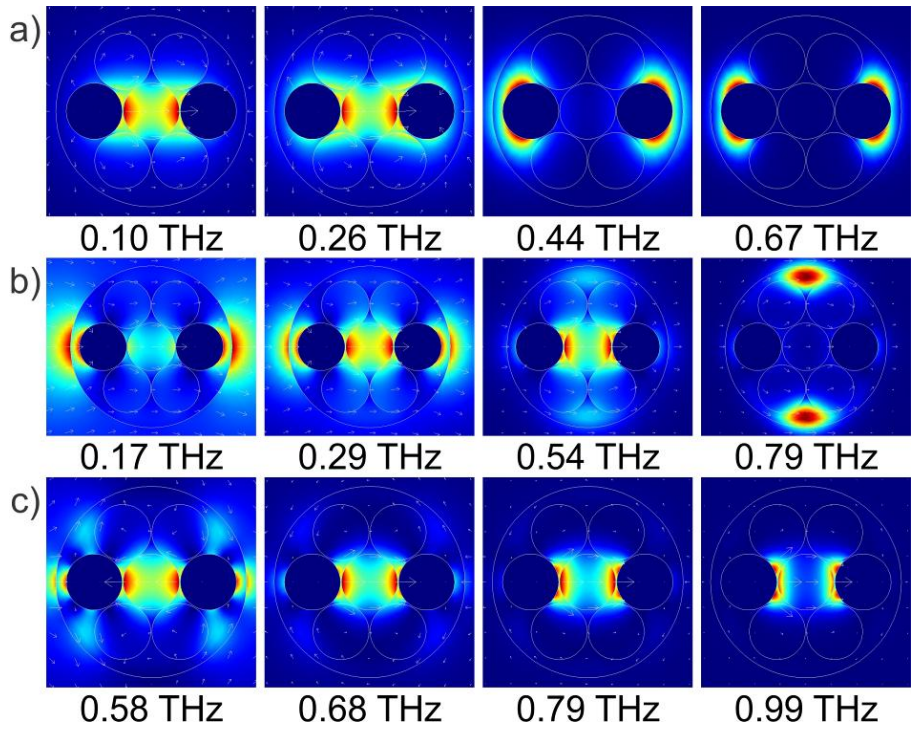

Fig. 13. Longitudinal flux distribution for the modes of a seven-hole composite fiber a) fundamental plasmonic mode, b) lowest order cladding mode, c) second plasmonic mode.

.

**Conclusion**

A novel type of practical THz fibers is proposed that combines low-loss, low-dispersion and efficient excitation properties of the classic two-wire waveguides together with mechanical robustness, and ease of manipulation of the porous dielectric fibers. We then show that while optical properties of composite fibers are inferior to those of a classic two-wire waveguide, at the same time, composite fibers outperform porous fibers of the same geometry both in bandwidth of operation and in lower dispersion. Finally, we demonstrated that by increasing porosity of the fiber dielectric cladding its optical properties could be consistently improved.

Particularly, using the finite element method we first ascertained that a classic two-wire waveguide features very low loss (<0.01 cm$^{-1}$) and very low group velocity dispersion (<0.1 ps/(THz·cm)) in the whole THz spectral range. We then studied coupling efficiency into such a waveguide as a function of various geometrical parameters and concluded that for an optimised waveguide, excitation efficiency of the fundamental mode can be relatively high (>50%) in the broad frequency range ~1 THz. Moreover, we confirmed that this excitation efficiency has a weak dependence on the misalignment in the position of a Gaussian excitation beam that was considered as the excitation source.

Next, we proposed using porous polyethylene cladding as a mechanical support for the two metal wires in order to provide a practical packaging solution for the classical two-wire waveguide. In their simplest implementation, resultant composite fibers feature three adjacent air-holes placed in a plastic cladding. Two peripheral holes are occupied by the metal wires, while the central one is used to guide the THz light. We then concluded that in a three-hole fiber the lowest order modes could be classified as either plasmonic modes or the modes of a porous cladding. This identification is possible when comparing field distributions and dispersion relations of the composite fiber modes with those of a corresponding porous fiber without metal wires, as well as with those of a classic two-wire waveguide.

Notably, the fundamental plasmonic mode of a composite fiber extends into very low frequencies (<0.1THz), while being well confined within the fiber. Both the fundamental and second plasmonic modes have reasonable excitation efficiencies of >10% in the broad frequency range 0.25 – 1.25 THz, while having relatively low group velocity dispersion of 1-3 ps/(THz·cm) and absorption losses that are, generally, 1.5–3 times smaller than the bulk absorption loss of a polyethylene cladding in the broad THz range 0.1 – 0.6 THz.

Similarly, the lowest order cladding mode of a composite fiber has, generally, lower absorption losses compared to those of the fundamental mode of the corresponding porous fiber without wires. At the same time, group velocity dispersion of this mode at low frequencies <0.3THz can be as high as 10 ps/(THz·cm), which is typical for porous fibers. Finally, coupling into the lowest order cladding mode of a composite fiber is quite efficient at lower frequencies < 1.0 THz and could be consistently above 50%.

We, therefore, conclude that composite three-hole fiber can somewhat outperform the corresponding porous fiber without wires, in terms of the bandwidth, absorption losses and lower group velocity dispersion if a restricted launch condition is used in order to preferentially excite the fiber plasmonic modes.

Finally, we demonstrated that the optical properties of a composite fiber could be consistently improved by further increasing the porosity of the fiber cladding. As an example, we considered a seven-hole composite fiber and demonstrated that such a fiber can have absorption losses that are at least ~3–5 times smaller than those of the bulk polyethylene in the broad THz frequency range 0.1 – 1.2 THz. Moreover, coupling efficiency into some plasmonic modes of such fibers was significantly improved (above 50% in the 0.75 – 1.25 THz range), thus enabling low-loss light transmission with low group velocity dispersion of <3 ps/(THz·cm). In this case of higher cladding porosity, the seven-hole composite fiber can clearly outperform the corresponding porous fiber both in the overall bandwidth and in the lower group velocity dispersion.